\documentclass[sigconf,natbib=true]{acmart}
\usepackage{multirow}
\usepackage{algorithm}
\usepackage{algorithmicx}
\usepackage{algpseudocode}
\usepackage{enumitem}  % 必须加载这个包才能使用 leftmargin=* 等参数
\usepackage{subfigure} 
\AtBeginDocument{%
  }

\acmISBN{978-1-4503-XXXX-X/2025/06}

\begin{document}

\title{Enhancing LLM-based Recommendation with Preference Hint Discovery from Knowledge Graph}
\author{
  Yuting Zhang$^1$,
  Ziliang Pei$^2$,
  Chao Wang$^3$,
  Ying Sun*$^1$,
  Fuzhen Zhuang$^4$
}
\affiliation{
  \institution{$^1$Thurst of Artificial Intelligence, The Hong Kong University of Science and Technology (Guangzhou),\\
  $^2$Institute of Computing Technology, 
   Chinese Academy of Sciences,\\ $^3$ University of Science and Technology of China, \\ $^4$ Institute of Artificial Intelligence, Beihang University}\country{}
}

% \author{Yuting Zhang}
% \affiliation{%
%   \institution{Thurst of Artificial Intelligence, The Hong Kong University of Science and Technology (Guangzhou)}
%   \city{Guangzhou}
% \email{yzhang755@connect.hkust-gz.edu.cn}

% \author{Ziliang Pei}
% \affiliation{
%   \institution{Institute of Computing Technology, 
%   Chinese Academy of Sciences}
%   \city{Beijing}
% \email{397538067@qq.com}

% \author{Chao Wang}
% \affiliation{
%   \institution{School of Artificial Intelligence and Data Science, University of Science and Technology of China}
%   \city{Heifei}
% \email{wangchaoai@ustc.edu.cn}

% \author{Ying Sun}
% \affiliation{
%   \institution{Thurst of Artificial Intelligence, The Hong Kong University of Science and Technology (Guangzhou)}
%   \city{Guangzhou}
% \email{yings@hkust-gz.edu.cn}

% \renewcommand{\shortauthors}{Yuting Zhang et al.}

\begin{abstract}
LLMs have garnered substantial attention in recommendation systems. Yet they fall short of traditional recommenders when capturing complex preference patterns. Recent works have tried integrating traditional recommendation embeddings into LLMs to resolve this issue, yet a core gap persists between their continuous embedding and discrete semantic spaces. Intuitively, textual attributes derived from interactions can serve as critical preference rationales for LLMs’ recommendation logic.  However, directly inputting such attribute knowledge presents two core challenges: (1) \textbf{Deficiency of sparse interactions in reflecting preference hints for unseen items};
(2) \textbf{Substantial noise introduction from treating all attributes as hints}. To this end, we propose a preference hint discovery model based on the interaction-integrated knowledge graph, enhancing LLM-based recommendation. It utilizes traditional recommendation principles to selectively extract crucial attributes as hints. Specifically, we design a collaborative preference hint extraction schema, which utilizes semantic knowledge from similar users’ explicit interactions as hints for unseen items. Furthermore,  we develop an instance-wise dual-attention mechanism to quantify the preference credibility of candidate attributes, identifying hints specific to each unseen item. Using these item- and user-based hints, we adopt a flattened hint organization method to shorten input length and feed the textual hint information to the LLM for commonsense reasoning. Extensive experiments on both pair-wise and list-wise recommendation tasks verify the effectiveness of our proposed framework, indicating an average relative improvement of over 3.02\% against baselines.

% LLMs have generated considerable interest in recommendation systems but struggle to understand ever-growing item catalogs and complex user preferences. Intuitively, attributes from user-interacted items can provide valuable insights into user preferences, thereby enhancing LLM-based recommendations. However, the raw attribute sets often provide poor hints for recommendations as sparse interactions limit insights for unseen items, as well as exhaustive item attributes can overwhelm the LLM. To this end, we introduce a preference hint discovery-based framework for enhancing LLM-powered recommendation. This framework utilizes traditional recommendation principles to extract crucial hints from knowledge graph to infer textual preference for LLM. Specifically, this framework consists of three main modules: (1) Collaborative user discovery module, inspired by the traditional collaborative filtering principle, supplements user's potential preferences via similar users to alleviate the sparse interaction issue. (2) Instance-wise preference hint discovery module selects crucial attribute-level preference  hints under traditional interaction-guided supervision, rather than inputting all attributes. (3) Preference prompt translation module converts discovered hints into a textual input form for LLMs. Extensive experiments have been conducted to evaluate the effectiveness of our proposed framework, indicating an average relative improvement of over 3.02\% compared to baselines in both pair-wise and list-wise recommendation tasks.
\end{abstract}


% <concept_id>10002951.10003317.10003347.10003350</concept_id>
% <concept_desc>Information systems~Recommender systems</concept_desc>



\maketitle
\section{Introduction}
Aiming to assist users in discovering interesting items from massive options, recommendation systems~\cite{he2017neural,mao2021simplex,bobadilla2013recommender} have emerged as indispensable components of various online platforms. For instance, Collaborative Filtering (CF)~\cite{su2009survey}, a representative paradigm of traditional recommender systems, assumes that users with similar behavioral patterns tend to share similar item preferences. However, such traditional methods are inherently constrained by the lack of commonsense knowledge~\cite{zhao2024let} and robust reasoning capabilities. Recently, Large Language Models~\cite{brown2020language, touvron2023llama, zhao2023survey}, equipped with rich world knowledge and emergent reasoning power, have offered a promising new solution for recommendation tasks~\cite{bao2023tallrec,zhang2023recommendation,zhang2023collm,liao2024llara,yang2023large, chen2024softmax}.

Along this line, initial explorations typically convert user historical interactions into natural language and query LLMs for recommendations~\cite{bao2023tallrec,chen2024softmax,zhang2023recommendation}. 
While LLMs' general world knowledge can support some reasonable recommendations, complicated preference patterns (such as global-view CF-related ones~\cite{su2009survey}) are difficult to seize, confining these initial approaches to suboptimal performance.
Therefore, recent studies~\cite{zhang2023collm,liao2024llara,yang2023large} have attempted to integrate continuous user/item embedding representations from traditional models into LLM inputs. Yet a fundamental gap exists between the continuous embedding space of traditional models and the discrete semantic space where LLMs operate. This \textbf{inherent semantic mismatch} impedes LLMs from fully exploiting their interpretable reasoning strengths for recommendations.

% Indeed, traditional recommendation methods are equipped with global-view recommendation logics tailored for inferring such nuanced preferences. For instance, user-based Collaborative Filtering (CF)~\cite{su2009survey} relies on the core assumption that users with similar behavioral patterns tend to share consistent item preferences.语义理解能力提供了新的可能。For instance, classical Collaborative Filtering (CF)~\cite{su2009survey} assumes that users with similar behavioral patterns tend to share consistent item preferences.. but 说一下缺点， 引出下一句对大语言模型的需求 %Recently, the extensive general knowledge and emergent reasoning capabilities of Large Language Models~\cite{brown2020language, touvron2023llama, zhao2023survey} have driven growing research on LLM-based recommendation systems~\cite{bao2023tallrec,zhang2023recommendation,zhang2023collm,liao2024llara,yang2023large, chen2024softmax}.  %While LLM's general world knowledge can support some reasonable recommendation, complicated user preference patterns implied in the entire recommendation platforms is difficult to be seized, confining LLMs to suboptimal performance. Yet relying merely on superficial interaction descriptions,,  LLM despite their general knowledge struggle to capture complex user preferences in real-world platforms

\begin{figure}
    \centering
    \includegraphics[width=0.8\linewidth]{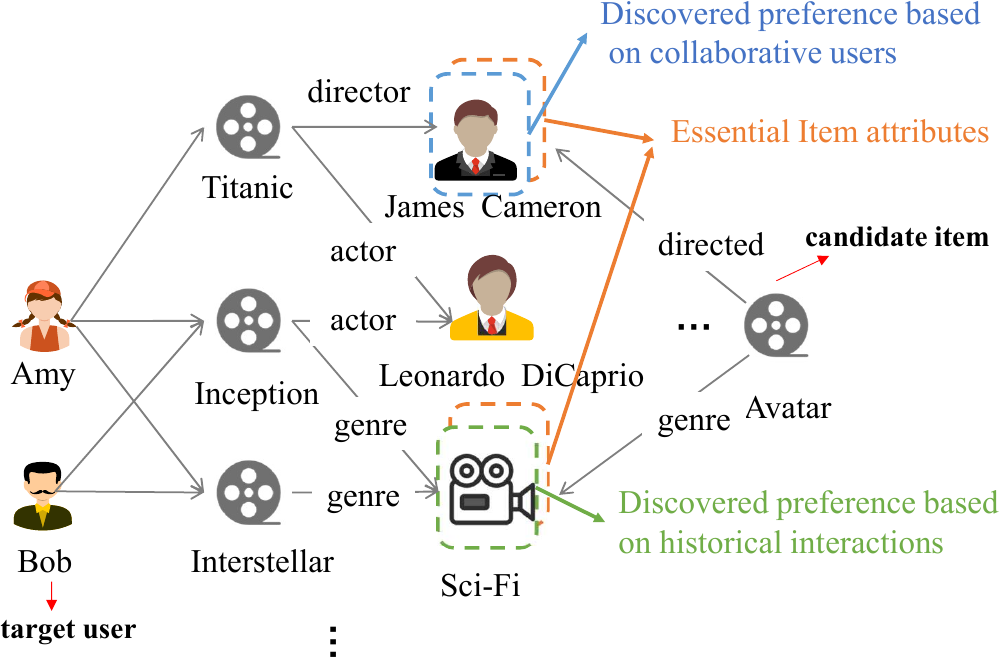}
    \caption{An illustration of preference hints discovered from interaction-integrated Knowledge Graph.}
    \label{fig:kg_rec}
\end{figure}

Intuitively, semantic attributes of user-interacted items inherently imply user preferences, enabling LLMs to leverage this knowledge for recommendation logic. For instance, the interaction attribute knowledge in Figure~\ref{fig:kg_rec} reveals that Bob interacts with movies \textit{Interstellar} and \textit{Inception} due to his preference for the \textit{Sci Fi} genre. Explicitly incorporating this preference hint into an LLM could thus enhance its ability to recommend the same-genre film \textit{Avatar} to Bob. Therefore, we propose to extract preference hints from \textbf{user-item interaction-integrated knowledge graph}  and feed such hints in textual form into LLM to boost recommendation performance. Nevertheless, this is non-trivial owing to the following challenges:
%Despite their extensive general knowledge, LLMs lack much item-specific attribute knowledge in pretraining corpora due to the rapidly evolving catalogs of real-world platforms (e.g., e-commerce~\cite{hussien2021recommendation}, short-video platforms~\cite{gong2022real}), thus hindering their recognition of user preference hints.

(1) \textbf{Sparse user interactions fail to reflect preference hints for unseen items}. To enable LLMs to understand user preference hints, one straightforward approach is to input all historical interaction attributes (e.g., \textit{Leonardo DiCaprio} and \textit{Sci-Fi} for user Bob, as shown in Figure~\ref{fig:kg_rec}). 
Nevertheless, the inherent sparsity of user interactions limits this approach’s ability to capture users' potential preferences for unseen items. Take Bob as an example: He might have an underlying interest in director \textit{James Cameron} of the unseen film \textit{Titanic}, but his sparse interaction attribute records fail to signal this preference. Consequently, this approach cannot enable LLMs to reason about such latent preferences or accurately recommend unseen items.

%Luckily, based on traditional CF principles, Bob might have a potential interest in director "James Cameron" due to similar interaction patterns with Amy. Yet traditional knowledge-aware recommenders indicate this potential via \textbf{\textit{non-semantic}} latent vectors, inherently incompatible with the semantic space of LLMs. Yet, due to interaction sparsity, this potential interest cannot be adequately captured by historical interaction attributes.

(2)  \textbf{Treating all item attributes as recommendation hints introduces noise}. Items are typically associated with multiple attributes; for instance, the widely used \textit{MovieLens} dataset~\cite{harper2015MovieLens} reports an average of 66 attributes for each movie.  Furthermore, different users tend to prioritize distinct attributes of even the same item, which indicates that critical attribute knowledge is inherently specific to each user-item instance (i.e., \textbf{\textit{instance-specific}}).
Therefore, feeding all available attribute knowledge into LLMs not only results in lengthier input texts but also, more critically, introduces substantial noise. Pretrained LLMs will be overwhelmed by the massive information and struggle to capture effective recommendation rationales from longer contexts.
%, especically from longer contexts~\cite{liu2024lost}. 

To address these challenges, we propose a novel \underline{P}reference h\underline{I}nt \underline{D}iscovery-based \underline{L}LM \underline{R}ecommendation enhancement framework (PIDLR). PIDLR utilizes traditional recommendation principles to selectively extract crucial  attributes from the user-item knowledge graph as preference hints to enhance LLM recommendation. Specifically, we propose a collaborative preference hint extraction schema, which leverages semantic knowledge from the explicit interactions of similar users as hints for unseen items. Furthermore, to reduce noise, we design an instance-wise dual-attention mechanism that quantifies the preference credibility of candidate attributes, identifying preference hints specific to each unseen item. Accordingly, based on these item- and user-based hints, we directly provide textual information to the LLM to fully leverage its commonsense reasoning capabilities. In particular, we incorporate a flattened textual hint organization method that effectively reduces the LLM’s input length. Extensive experiments conducted for pair-wise and list-wise recommendation tasks on two publicly available recommendation datasets demonstrate the superiority of our model.

\section{Related work}\label{sec:related work}
\subsection{LLM-based Recommedantion}
Driven by the rapid advancement of LLM, researchers have explored integrating LLMs into recommender systems (termed LLM4Rec~\cite{wu2024survey}), which falls into two core categories:
(1) \textbf{LLM as an Enhancer}.
LLMs~\cite{ren2024representation,geng2024breaking,xi2024towards} act as feature extractors or text generators to augment traditional recommendation algorithms by enriching user/item profiles with multi-source data (e.g., item metadata). However, the core recommendation pipeline remains conventional, thus failing to fully exploit LLMs’ reasoning capabilities.
(2) \textbf{LLM as a Recommender}.  Early work utilized LLMs’ in-context learning for zero/few-shot recommendations~\cite{hou2024large,liu2023chatgpt}. Subsequent efforts optimized LLMs via supervised fine-tuning on historical interaction data, and recent studies show that incorporating traditional recommendation system information further enhances LLM-based recommendation performance~\cite{zhang2023collm, liao2024llara,yang2023large}. However, these methods directly adopt id-level user/item embeddings from traditional models. Such embeddings lack semantic richness and mismatch LLMs’ semantic space, thus compromising efficacy.
In contrast, our approach first extracts semantic preference hints based on traditional recommendation principles and integrates them into LLMs. This design enables LLMs to better grasp preference logic, thereby improving recommendation performance. % enabling better understanding of preference logic and improved recommendation performance.

\subsection{Knowledge-aware Recommendation}
Traditional knowledge-aware recommendation methods fall into three categories~\cite{guo2020survey}:
(1) Embedding-based methods~\cite{zhang2016collaborative,cao2019unifying}: Enrich user or item representations with knowledge graph (KG) entities/relations (e.g., CKE~\cite{zhang2016collaborative} fuses side information via TransR~\cite{lin2015learning}). (2) Path-based methods~\cite{hu2018leveraging,HAN}: Explore KGs to uncover long-range connections between entities, which can identify more complex relationships beyond direct connections.
%Mine long-range entity connections in KGs. 
(3) GNN-based methods~\cite{wang2019kgat,yang2022knowledge}: Aggregate multi-hop neighbor embeddings via GNNs to capture rich structural information of KG. However, these methods map textual entities/relations to IDs, limiting the recommendation system’s ability to fully leverage the KG’s potential. Recent work integrates LLMs with KGs for recommendation: (1) Studies like CoLaKG~\cite{cui2024comprehending} and LLMRG~\cite{wang2024llmrg} use LLMs to generate semantic embeddings or reasoning graphs for traditional recommendation enhancement. (2) Works such as KGLA~\cite{guo2024knowledge} and KGGLM~\cite{balloccu2024kgglm} feed KG meta-paths into LLMs for direct recommendation. However, the inherent interaction sparsity limits these paths' ability to reflect users' comprehensive preferences.
Unlike prior approaches, our method proposes to leverage semantic knowledge from similar users’ explicit interactions as hints for unseen items, enabling comprehensive preference learning.

\section{Preliminaries}\label{sec:preliminary}
Let $\mathcal{U}$ and $\mathcal{V}$ denote the universal sets of users and items, respectively. In a recommendation scenario, each user $u \in \mathcal{U}$ has chronologically interacted with an item sequence $B_u = [v^u_1, v^u_2, \cdots, v^u_n]$, where $v^u_i \in \mathcal{V}$.
To capture additional attribute knowledge, we introduce a heterogeneous graph as follows: %It provides valuable supplementary information to enhance item characteristic and relationship understanding

% \textbf{Interaction-integrated Knowledge Graph}. 
% A heterogeneous graph is composed of triplets, denoted as $\mathcal{G}_k = \{(h, r, t) \mid h, t \in \mathcal{E} \land r \in \mathcal{R}\}$. Here, $\mathcal{E}$ represents the set of entities, which encompasses both the universal user set $\mathcal{U}$ and item set $\mathcal{V}$ (i.e., $\mathcal{U} \cup \mathcal{V} \subset \mathcal{E}$). $\mathcal{R}$ denotes the set of relations, including not only the inherent semantic relations between entities (e.g., attribute, affiliation) but also the user-item interaction relations derived from recommendation scenarios. Specifically, $h$ and $t$ are the head and tail entities, respectively, while $r$ is the semantic relation connecting them. For example, the triplet \textit{(Titanic, director, James Cameron)} states the fact that \textit{James Cameron} is the \textit{director} of the movie \textit{Titanic}; another typical triplet, such as \textit{(Bob, interacted, Titanic)}, which records the interaction behavior of a user with the item. %Notably, the item set $\mathcal{V}$ and user set $\mathcal{U}$ are both subsets of the entity set $\mathcal{E}$, denoted as $\mathcal{U} \subset \mathcal{E}$ and $\mathcal{V} \subset \mathcal{E}$.

\noindent\textbf{Knowledge Graph}. A heterogeneous graph is composed of triplets, denoted as $\mathcal{G}_k = \{(h, r, t) \mid h, t \in \mathcal{E} \land r \in \mathcal{R}\}$. Here, $\mathcal{E}$ represents the set of entities, and $\mathcal{R}$ represents the set of relations. Specifically, $h$ and $t$ are the head and tail entities, respectively, while $r$ is the semantic relation connecting them. For example, the triplet \textit{(Titanic, director, James Cameron}) states the fact that \textit{James Cameron} is the \textit{director} of the movie \textit{Titanic}. Notably, the item set $\mathcal{V}$ is a subset of the entity set $\mathcal{E}$, denoted as $\mathcal{V} \subset \mathcal{E}$.

With interaction-integrated knowledge graph, the recommendation task can be defined as follows:

\noindent \textbf{Task Formulation}. Given historical interactions $B_u$ of user $u$, and knowledge graph $\mathcal{G}_k$, our objective is to predict the next item $v$ preferred by user $u$ from a candidate set $V = \{v_i\}_{i=1}^m$, where $m$ is the number of candidates. Each pair $(u, V)$ constitutes a recommendation \textit{\textbf{instance}}.

This task aims to leverage interaction patterns and knowledge graphs for accurate, reasoning-based recommendations. For each instance, we need to incorporate an instance-specific \textit{\textbf{preference hint}}, such as identifying that a user potentially favors \textit{director James Cameron} and that a candidate item shares similar semantic attributes.  Equipped with such preference hints, LLMs with robust reasoning capabilities can better grasp recommendation logic and generate rational recommendations.  Additionally, to further refine performance in this paradigm, instruction tuning is a crucial step~\cite{ouyang2022training} to boost task-specific instruction-following ability:

%based on those preference( the \textbf{preference hint} for the recommendation),  an aspect that LLMs with strong reasoning ability excel at. to help LLM understand the underlying recommendation logic
\noindent\textbf{Instruction Tuning}. LLMs are fine-tuned to better understand and execute task-specific instructions, using instruction-response pairs   $\mathcal{Z} =\{(x_i,y_i)\}_{i=1}^M$ ($x_i=$ textual instructions and $y_i=$ corresponding responses). For example: instruction "Do you like Titanic or Avatar?" and response \textit{Avatar}. Based on $\mathcal{Z}$,  LLMs are tuned via the following autoregressive objective:
\begin{equation}
\begin{aligned}
    \max _{\Theta} \sum_{(x, y) \in \mathcal{Z}} \sum_{t=1}^{|y|} \log (P_{\Theta}(y_t \mid x, y_{<t})),
\end{aligned}
\end{equation}
where $y_t$ denotes the $t$-th token of $y$, $y_{<t}$ represents the tokens before $y_t$, $\Theta$ is the original parameters of LLM. 

\section{Methodology}\label{sec:method}
\begin{figure*}
    \centering
    \includegraphics[width=0.95\linewidth]{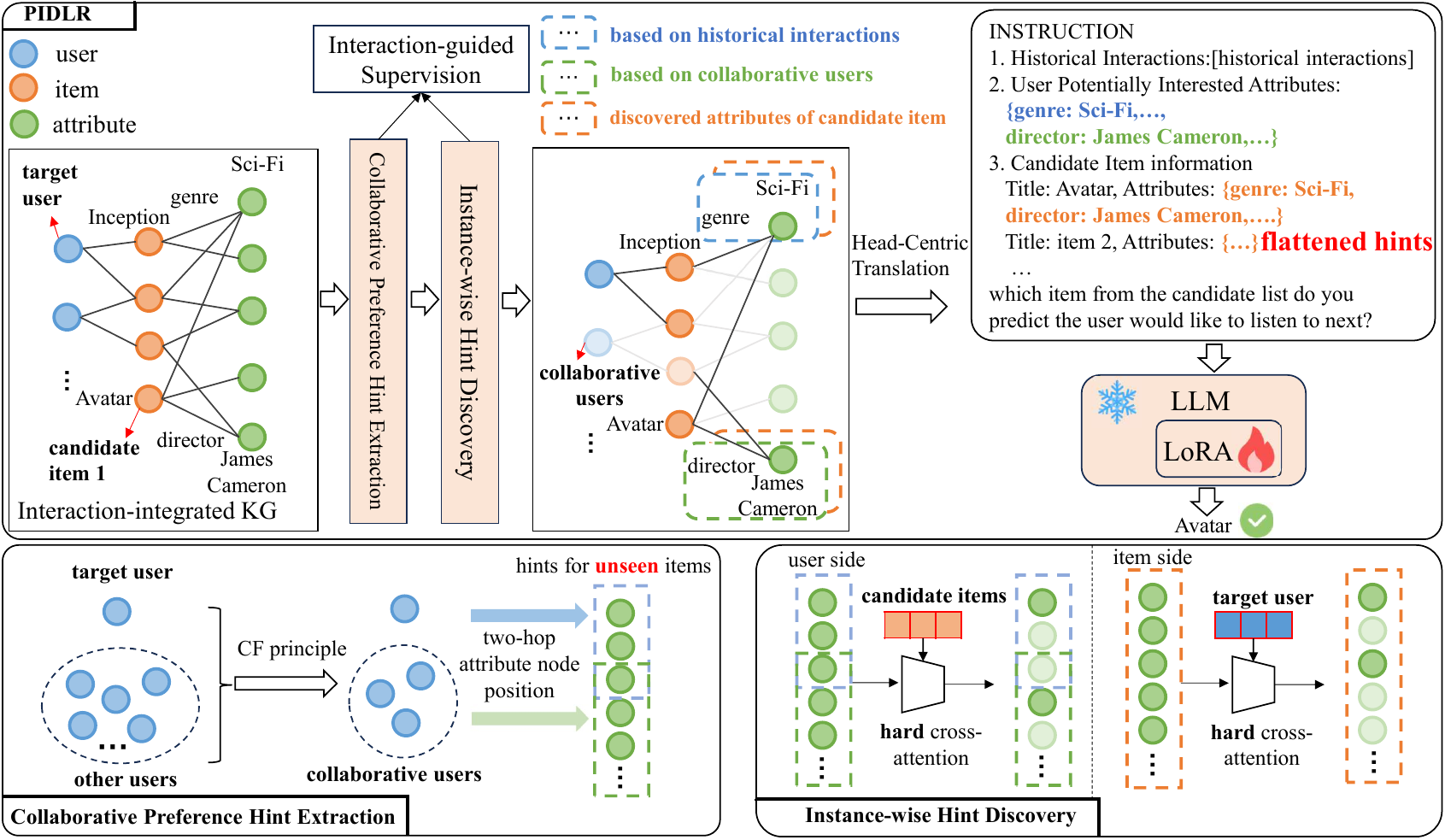}
    \caption{Overall Framework of PIDLR. Initially, the Collaborative Preference Hint Extraction module supplements users' potential preference hints  preference hints for unseen items based on collaborative filtering principle. Then, the Instance-wise Hint Discovery module identifies the personalized preference hints for each instance. Finally, Head-Centric Translation converts the identified preference hints into flattened texts for LLM-based recommendation.
    }
    \label{fig:framework}
\end{figure*}

%与传统LLM推荐的区别
In this section, we introduce PIDLR for discovering preference hints to enhance LLM-based recommendation, with its framework illustrated in Figure~\ref{fig:framework}.
First, Section~\ref{sec:coll_dis} details the Collaborative Preference Hint Extraction module, which mines user comprehensive preferences via collaborative signals. Next, Section~\ref{sec:hint discovery} presents the Instance-wise Hint Discovery module, which designs a dual-attention mechanism to identify critical preference hints for each recommendation instance. Then, Section~\ref{sec:prompt_translator} describes the Head-Centric Hint Translation module, which converts discovered hints into flattened textual prompts to shorten the input length. Finally, translated prompts are fed into the LLM to drive the interpretable, hint-based recommendations.

\subsection{Collaborative Preference Hint Extraction}\label{sec:coll_dis}

User interaction sparsity makes sole reliance on historical interaction data for LLM input insufficient to capture comprehensive user preferences. To address this, we propose a Collaborative Preference Hint Extraction module inspired by traditional user-based collaborative filtering principles.
This module’s core idea is to calculate user similarity from historical interaction data and infer comprehensive user preferences using similar users' preferences, as detailed below. %%It aims to leverages semantic knowledge from the explicit interactions of similar users as hints for unseen items.  it calculates user similarity via historical interaction preferences and predicts target user interests based on similar users' preferences. Thus,

First, we collect each user $u$’s historical interactions as $B_u=[v_u^1,v_u^2,\cdots,v_u^n]$.  Then, for each item $v \in B_u$, we extract its first-order attribute subgraph (ego network~\cite{qiu2018deepinf}) from the knowledge graph $\mathcal{G}_k$, represented as $\Gamma_v=\{(r,t)|(v,r,t)\in \mathcal{G}_k\}$. On this basis, the two-hop attribute subgraph centered on $u$ is the union of all $\Gamma_v$ for $v \in B_u$, denoted as $\Gamma_u = \cup_{v\in B_u} \Gamma_v$. This subgraph includes $u$’s interacted items and these items' one-hop attributes. We then obtain this subgraph’s representation as follows.

%providing dual views of $u$’s item-level and attribute-level preferences. 

Let $e_u \in \mathbb{R}^d$ and $e_v \in \mathbb{R}^d$ denote the ID embeddings of user $u$ and item $v$ respectively, with $d$ denoting the dimension. For each tuple $k\in \Gamma_v$, we reindex all tuple IDs and obtain their corresponding embeddings, denoted as $e_k$. The representation of user-centered subgraph $\Gamma_u$ is then formulated as:
\begin{equation}\label{eq:user_representation}
\begin{aligned}
    E_u = e_u\oplus mean(\{e_v,v\in B_u\})\oplus mean(\{e_k,k\in\Gamma_u\}),
\end{aligned} 
\end{equation}
where $\oplus$ denotes the concatenation operation and $mean$ denotes the mean pooling operation widely used in graph aggregation~\cite{hamilton2017inductive}. 
Then, using user representation $E_u$, our module calculates the similarity between $u$ and other users $u'$.  
For efficiency, we take same-batch users as candidate set $\mathcal{U}_b$ and identify the top-N via similarity evaluation across $\mathcal{U}_b$, as follows:
\begin{equation}\label{eq:col_num}
    \begin{aligned}
        U = \arg \text{TopN}_{u'\in \mathcal{U}_b \textbackslash u} cosine(\textbf{W}_cE_u,\textbf{W}'_cE_{u'}),
    \end{aligned}
\end{equation}
where $\textbf{W}_c$, $\textbf{W}'_c$ denote linear transformations and $cosine(\cdot,\cdot)$ is a widely used similarity measure. In this work, users $u'$ in $\mathcal{U}$ are referred to as \textbf{collaborative users} to distinguish them from similar users identified via user features (e.g., age, gender). After identifying collaborative users, we infer that $u$ shares similar preferences with $u' \in \mathcal{U}$ based on collaborative filtering principles. We then locate these collaborative users $u'$ in the knowledge graph and obtain their user-centered two-hop attribute subgraphs (i.e., $\Gamma_{u'}$) to infer $u$'s comprehensive preferences for unseen items.
\subsection{Instance-wise Hint Discovery}\label{sec:hint discovery}

%Traditional item-based collaborative filtering~\cite{su2009survey} assumes users will interact with items similar to previously interacted ones. In this study, we use item attributes to measure similarity, leading to the assumption that users prefer candidate items whose attributes closely align with their attribute-level preferences. 

% Consequently, both candidate item attributes and user attribute-level preferences can serve as hints to assist LLMs in understanding recommendations. To implement this, user-preferred attributes and item attributes need to be input. However, treating all attributes in $\hat{\Gamma}_u$ as user preference hints introduces significant noise and 

Since treating all attributes of each instance as recommendation hints introduces noise, this module aims to select instance-specific hints from the comprehensive potential preferences $\hat{\Gamma}_u$ and candidate item attributes $\Gamma_v$. Specifically, it adopts a symmetric dual-attention mechanism to simultaneously identify key attributes from both the user’s comprehensive potential attribute preferences $\hat{\Gamma}_u$ and candidate item attributes $\Gamma_v$. This dual process integrates user preference discovery and item attribute discovery, as detailed below. %, ensuring a more tailored recommendation process for each instance.

\subsubsection{User Preference Discovery}\label{sec:User_Preference_Discovery}
In Section~\ref{sec:coll_dis}, we obtained a comprehensive set of potential preferences $\hat{\Gamma}_u$ for user $u$, consisting of attributes from $u$’s historically interacted items and collaborative users $u'$.
Treating all such attributes as preferences introduces substantial noise, and informative preference hints vary across candidate item sets. To address these issues, we propose a User Preference Discovery module that extracts instance-aware informative preferences from $\hat{\Gamma}_u$ via selective augmentation. Specifically, given the candidate item set $V$ in the instance(serving as Key vector in Cross-Attention mechanism), we calculate the preference credibility score of each attribute in $\hat{\Gamma}_u$ (serving as Query vector), defined as follows:

%This process identifies the relevant knowledge that best aligns with the user's preferences for each specific instance as follows:

\begin{equation}\label{eq:user_attr_p}
    \begin{aligned}
     \forall k \in \hat{\Gamma}_u,\quad p(k|(u,V))=\frac{{\rm exp}( \textbf{W}_u\mathbf{E}_{V} \times {\textbf{W}'_u\mathbf{e}_{k}^\top})}{\sum_{i \in \hat{\Gamma}_u}{\rm exp}( \textbf{W}_u\mathbf{E}_{V} \times {\textbf{W}'_u\mathbf{e}_{i}^\top)}}, \\
    where \quad \mathbf{E}_{V} = \oplus_{v \in V} (mean(\{e_j,\forall j\in\Gamma_v\})),
    \end{aligned}
\end{equation}
where $\textbf{W}_u$, $\textbf{W}'_u$ denote linear transformations for aligning dimensions of $\mathbf{E}_{V}$ (attribute-level representation of candidate set $V$) and $\mathbf{e}_{k}$. The preference credibility  quantifies the relevance of each attribute in $\hat{\Gamma}_u$ to item-side attribute information; higher scores indicate stronger preference credibility with the candidate set and greater likelihood of reflecting the user preferences.
To identify key attributes, we adopt a sampling strategy based on these scores: sampling $\alpha_u *|\hat{\Gamma}_u|$ attributes from $\hat{\Gamma}_u$ via $p(k|(u,V))$. Parameter $\alpha_u<1$ controls the selection proportion, yielding $\widetilde{\Gamma}_u \subset \hat{\Gamma}_u$. This study designs a \textbf{hard} cross-attention mechanism, which assigns a weight of 0 to attributes with low credibility $p(k|(u,V))$ and a weight of 1 to those with top credibility. Unlike soft attention schemes that allocate soft weights to all attributes, this hard mechanism directly selects a subset of attributes to fulfill the selection objective. This subset highlights the most informative attributes relevant to the user’s preferences in the context of $V$, enabling LLM to accurately capture instance-specific preferences.% thereby enhancing the precision and relevance of the recommendation results.
%This subset is specifically chosen to best illuminate the user's attribute-preference learning. 

\subsubsection{Item Attribute Discovery}\label{sec:item_Attribute_Discovery}
For each candidate item $v \in V$, we extract its one-hop attribute subgraph $\Gamma_v$ (reflecting the direct attributes of $v$) from the knowledge graph $\mathcal{G}_k$. However, the attribute sets of $v \in V$ are often extensive (e.g., MovieLens items have an average of 66 attributes). While informative, this extensiveness poses challenges for LLMs: prior work~\cite{liu2024lost} shows LLMs struggle to capture key factors from lengthy contexts. Since different users focus on different attribute subsets of the same item, the Item Attribute Discovery module is designed to selectively extract instance-specific significant hints. Similar to the hard attention in the User Preference Discovery module, given user $u$ (serving as the key vector) in the instance, we symmetrically calculate the preference credibility score of each attribute in $\Gamma_v$, as follows:

\begin{equation}\label{eq:item_attr_p}
    \begin{aligned}
    \forall k \in \Gamma_v,\quad p(k|(u,V)) = \frac{{\rm exp}(\textbf{W}_v\mathbf{E}_{u} \times {\textbf{W}_v\mathbf{e}_{k}^\top)}}{\sum_{i \in \Gamma_v}{\rm exp}(\textbf{W}_v\mathbf{E}_{u} \times {\textbf{W}'_v\mathbf{e}_{i}^\top)}}, 
    \end{aligned}
\end{equation}
Next, based on their preference credibility s, we also sample $a_v*|\Gamma_v|$ attributes from $\Gamma_v$ for each item $v \in V$. This sampled subset is denoted as $\widetilde{\Gamma}_v$. By focusing on the most significant attributes, $\widetilde{\Gamma}_v$ can make the subsequent LLM-based recommendation process both efficient and effective, leveraging the most significant information for each instance to generate high-quality recommendations.

\subsubsection{Instance Interaction-guided Supervision}
To enable attributes $\widetilde{\Gamma}_u$ (user-side preference hints) and $\widetilde{\Gamma}_v$ (item-side  preference hints) to reflect the interaction rationales accurately, we adopt a traditional interaction supervised learning method. It uses the ground-truth interaction signal to end-to-end guide the learning of preference hints derived from the Collaborative Preference Hint Extraction and Instance-wise Hint Discovery modules. Specifically, for each candidate item $v \in V$ in each instance, we concatenate $\widetilde{\Gamma}_u$ with the corresponding $\widetilde{\Gamma}_v$ of the candidate item, and a vanilla MLP takes these concatenated features as input to predict the preference scores of all candidate items in $V$, formulated as follows:
\begin{equation}
    \begin{aligned}\label{eq:predict}
    \hat{y}_{u,v}&={\rm MLP}(\widetilde{\mathbf{E}}_{u}\oplus \widetilde{\mathbf{E}}_{v}), \\
    where \quad & \widetilde{\mathbf{E}}_{u}=\oplus_{k\in \widetilde{\Gamma}_u} e_{k}, \quad \widetilde{\mathbf{E}}_{v}=\oplus_{k\in \widetilde{\Gamma}_v} e_{k}.
    \end{aligned}
\end{equation}

We then adopt the widely used Bayesian Personalized Ranking (BPR) loss~\cite{rendle2009bpr} to enforce higher scores for positive than negative interactions, as follows:
\begin{equation}\label{eq:bpr_loss}
    \begin{aligned}
        \mathcal{L}_{o}=-\sum\limits_{(u,V)\in \mathcal{D}}\sum_{v^-\in V}\ln \sigma(\hat{y}_{u,v^+}-\hat{y}_{u,v^-}),
    \end{aligned}
\end{equation}
where $\mathcal{D}$ denotes the training dataset; $\sigma(\cdot)$ is the sigmoid function; $v^+$ denotes observed interactions, and $v^-$ denotes randomly sampled unobserved items from the candidate set $V$.
We then optimize objective function $\mathcal{L}_{o}$ to learn parameters of the collaborative preference hint extraction and instance-wise hint discovery modules. After parameter convergence, we identify the key attribute hints $\widetilde{\Gamma}_u$ and $\widetilde{\Gamma}_v$ for all instances in training and testing datasets.
% The detailed algorithm for discovering preference hints, which combines the instance-wise hint discovery and collaborative preference hint extraction modules, is presented in Algorithm \ref{alg:PreferenceHint}.

% The knowledge graph provides additional semantic information about items and their attributes. This external knowledge can supplement the limited interaction data, helping the model to better understand the context and relationships between items and attributes.

\subsection{Head-Centric Hint Translation}\label{sec:prompt_translator}
%Since the input to LLM is in text format, our objective is to convert  the preference-guided selected knowledge and collaborative information into a textual representation that the LLM can comprehend.
% For LLM recommendation, knowledge information should also be represented by text to effectively influence LLM recommendation process.

After obtaining the key instance-wise attribute hints $\widetilde{\Gamma}_u$ and $\widetilde{\Gamma}_v$, we further convert these hints into LLM-interpretable textual representations.
Existing knowledge representation methods for LLMs typically adopt a triple-based format~\cite{lisimple}. However, in recommendation scenarios, the head entities (i.e., users or items) are shared across all triples within the same instance-wise hint set. To address this redundancy, we propose a Head-Centric Hint Translation strategy, which retains each shared head entity only once instead of repeating it across multiple triples. This strategy organizes the associated semantic attributes of each head entity into a flattened attribute set, which serves as the knowledge prompt for LLMs. This design reduces the length of knowledge prompts by nearly one-third, thereby mitigating the computational overhead caused by redundant head entity encoding while preserving all semantic attributes required for recommendation tasks. Specifically, the prompt translation process consists of two components: user-centric hint prompt transforms $\widetilde{\Gamma}_u$ into coherent textual descriptions that capture instance-specific user preferences; item-centric hint prompt converts the core item attributes $\widetilde{\Gamma}_v$ into text that delineates the key instance-wise attributes of candidate items.
 %In the following, we translate user-side and item-side hints into text format, respectively.

\subsubsection{User-centric Hint Prompt}
First, we incorporate the attribute preference hints $\widetilde{\Gamma}_u$ of user $u$ into the prompt. Each element $k=(r,t)$ in $\widetilde{\Gamma}_u$ denotes an attribute tuple for the same user, consisting of relation $r$ and tail entity $t$. For example, the structure $\text{Text}(r):\text{Text}(t)$ can be instantiated as \textit{director: James Cameron}, indicating that user $u$ has a potential interest in movies directed by \textit{James Cameron}. The translation of $\widetilde{\Gamma}_u$ is formulated as follows:
\begin{equation}\label{eq:user_attr}
    P_c = \big\Vert_{\forall (r,t)\in \widetilde{\Gamma}_u}\{\text{Text}(r):\text{Text}(t)\},
\end{equation}
where $\big\Vert$ denotes the textual concatenation operation, which connects the text representations of different attributes using commas as separators. Besides, $\text{Text}(r)$ and $\text{Text}(t)$ refer to the textual names of relation $r$ and tail entity $t$, respectively. Thus, $P_c$ presents the potential instance-wise attribute preferences of user $u$ to the LLM in textual form.

Additionally, following existing LLM-based recommendation methods, we incorporate the user's historical interaction behavior into the prompt design. Specifically, we concatenate all items in the historical interaction set $B_u$ of user $u$ to construct an item-level preference prompt, defined as follows:
\begin{equation}\label{eq:interaction}
    P_b = \big\Vert_{v\in B_{u}} \text{Text}(v).
\end{equation}

Through the above flattened translation steps, we can simultaneously feed two types of user preferences, item-level preferences and attribute-level preferences, into the LLM, with both preference types centric to the same target user $u$.

\subsubsection{Item-centric Hint Prompt}
For each candidate item $v$ in $V$, 
we merge the relations and the tail entities in $\widetilde{\Gamma}_v$ to form a flattened attribute-aware text description of each head item $v$ as follows:
\begin{equation}\label{eq:item_attr}
    \begin{aligned}
    P_v = (Title:\text{Text}(v), Attributes:\{\Vert_{(r,t)\in \widetilde{\Gamma}_v} \text{Text}(r):\text{Text}(t)\}).
    \end{aligned}
\end{equation}

Compared to the previous works, the input prompt $P_v$ for each candidate item now additionally incorporates essential instance-aware attributes. For example, $P_v$ can be formatted as \textit{(Title: Avatar, Attributes: \{genre: science fiction, director: James Cameron,….\})}, including candidate item attributes that reflect instance-wise textual hints under interaction-based supervision.

\begin{figure}
    \centering
    \includegraphics[width=0.9\linewidth]{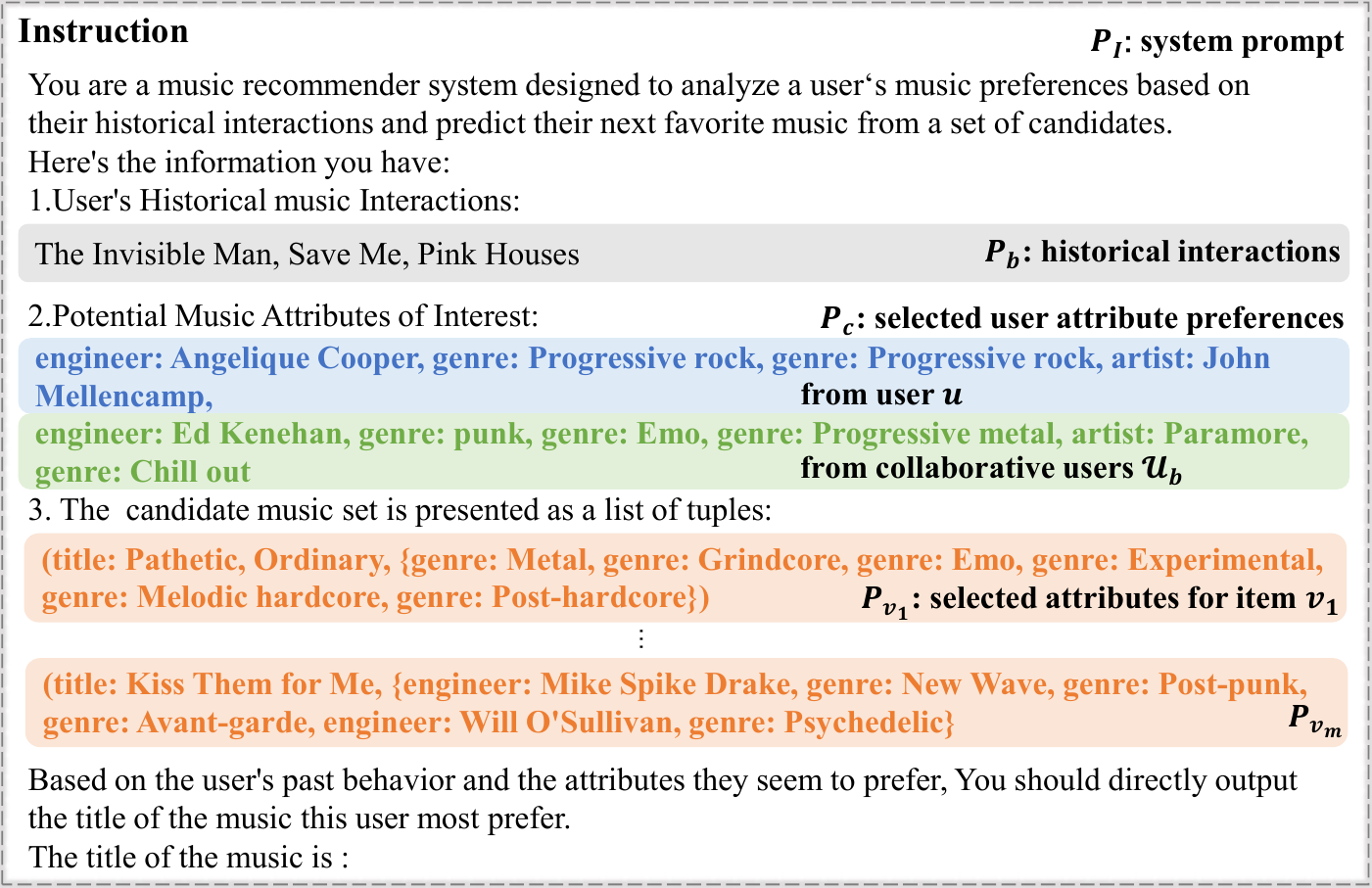}
    \caption{Prompt example on LastFM.}
    \label{fig:example_prompt}
\end{figure}

The aforementioned head-centric prompts $P_b$, $P_c$, $P_v$ retain each shared head entity only once instead of repeating it across multiple triples.  Also, this design alleviates the difficulty for LLM to comprehend that the same entity is associated with multiple semantic hints.  Besides the head-centric prompts, we design a system prompt $P_I$ to guide the LLM to follow the recommendation process. By combining $P_I$ with $P_b$, $P_c$, $P_v$, we construct the final prompt as:
\begin{equation}\label{eq:final_prompt}
    \begin{aligned}
        x = \Vert(P_I,P_b,P_c,\{P_v\}_{v\in V}).
    \end{aligned}
\end{equation}
The final prompt composition is illustrated in Figure~\ref{fig:example_prompt}. Notably, integrating the discovered attributes into the LLM serves two purposes: first, attribute selection reduces the input volume and simplifies the input; second, highlighting informative preference hints explicitly guides LLM toward more effective reasoning. This prompt enables the LLM to fully comprehend the informative, instance-specific recommendation knowledge.
 %: gray box presents user historical interactions, blue box highlights selected user attribute preferences, green box displays collaborative users’ selected attribute preferences, orange box indicates candidate item selected attributes. Remaining text is the system prompt.
\subsubsection{Fine-tuning process}
We then introduce instruction tuning and the recommendation process to efficiently align the LLM with the recommendation task. As full fine-tuning of LLM parameters is computationally expensive and time-consuming, we adopt the Parameter-Efficient Fine-Tuning (PEFT) algorithm LoRA~\cite{hu2021lora}. LoRA freezes the original LLM parameters and injects trainable low-rank matrices into each Transformer layer, enabling lightweight fine-tuning and reducing GPU memory consumption.
The optimization objective is formulated as:
\begin{equation}\label{eq:objective}
\begin{aligned}
    \max _{\Theta} \sum_{(x, y) \in \mathcal{Z}} \sum_{t=1}^{|y|} \log (P_{\Theta+ \Theta_L}(y_t \mid x, y_{<t})),
\end{aligned}
\end{equation}
where $\Theta_L$ denotes the LoRA parameters, which are significantly smaller in size than the original LLM parameters $\Theta$.
%The detailed algorithm for translating preference hints and fine-tuning the LLM using the translated preference hints as input is presented in Algorithm \ref{alg:prompt_translator}.

%and the  next item title of user
\section{Experiment}\label{sec:experiments}
In this section, we present empirical results to validate our proposed PIDLR, addressing the following research questions:
\textbf{RQ1} How does PIDLR perform compared with traditional recommender models and LLM-based methods? \textbf{RQ2} What are the effects of the collaborative preference hint extraction and instance-wise hint discovery modules in PIDLR? \textbf{RQ3} How do the preference hints discovered by PIDLR influence the LLM-based recommendation performance? \textbf{RQ4} How do the hyperparameters in PIDLR impact the recommendation performance? \textbf{RQ5} How does PIDLR perform across varying data sizes, especially in few-shot scenarios?

\subsection{Experimental Settings}

\subsubsection{Dataset Description}
To evaluate the effectiveness of PIDLR, we conducted extensive experiments on two widely-used datasets: (1) \textbf{MovieLens}~\cite{harper2015MovieLens}:  a well-known movie recommendation dataset with user ratings and movie titles. (2) \textbf{LastFM}~\cite{schedl2016lfm}: a music dataset consisting of users' listening interactions and music names.
For both datasets, we utilize the knowledge graph from~\cite{balloccu2024kgglm}. 
%and following the setting in~\cite{liao2024llara}. 
Dataset sizes are comparable to prior works~\cite{bao2023tallrec,liao2024llara}, ensuring similar experimental scale and complexity.
%For both datasets, we  partitioned the data into training, validation, and testing sets using an 8:1:1 ratio by timestamp. \textbf{The size of these datasets is comparable to those used in previous works}~\cite{bao2023tallrec,liao2024llara}, ensuring that our experiments are conducted on datasets of similar scale and complexity.
%Given the great computational demands of tuning LLMs, we randomly sample one-third of the users and items from each dataset, retaining their interactions to create a moderately-sized dataset following~\cite{liao2024llara}, 
Detailed dataset statistics are presented in Table~\ref{tab:data_info}. 
%We can observe that MovieLens dataset exhibits a higher density and average item degree compared to LastFM. In contrast, LastFM contains a larger number of items and interactions.  Additionally, MovieLens has a greater variety of entity types and relation types, whereas LastFM has a more extensive knowledge graph. 
Notably, the average item degree in both datasets exceeds 20, with MovieLens reaching 66.  This high connectivity underscores the need to select critical attribute information for recommendation tasks, especially list-wise ones. For evaluation consistency, we followed~\cite{liao2024llara}: retained the last 10 interactions as historical sequences and padded sequences shorter than 10 for uniformity.

\begin{table}[]\small
    \caption{Statistics of the datasets. 
    }
    \centering
    \begin{tabular}{c||c|rr}
    \toprule
    &Name& \textbf{MovieLens} & \textbf{LastFM}\\
    \midrule
    \multirow{4}{*}{\#Interaction}&User     &2,013 &1,604  \\
    &Item     &991  &4,164 \\
    &Interaction &100,169 &118,136 \\
    &Density &0.050 &0.017 \\
     \midrule
    \multirow{4}{*}{\#Knowledge} &Entity (Types) & 10,292(12)&7,387(5)\\
    &Relations (Types) &65,979(11)&76,425(4)\\
    &Avg. Degree (All) &12.822 & 18.358 \\
    &Avg. Degree (Item) &66.645 &20.692 \\
    \bottomrule
    \end{tabular}

    \label{tab:data_info}
\end{table}

\subsubsection{Evaluation Metrics} 
We design two classic recommendation tasks to comprehensively assess the performance and generalization ability of our model:
\begin{itemize}[leftmargin=*,itemsep=2pt,topsep=0pt,parsep=0pt]
    \item Pair-wise recommendation task. The task learns to distinguish user-preferred items from non-preferred ones. For each positive pair $(u,v^+)$, we randomly sample a non-interacted item $v^-$ to form a 2-item candidate set (i.e.,$|V|=2$).
    \item List-wise recommendation task. The task ranks candidate items to identify the top preference. Following~\cite{liao2024llara,gao2023chat}, we construct a 20-item candidate set $|V|=20$ for each instance by sampling 19 non-interacted items plus the ground-truth next item.
\end{itemize}
For both tasks, all models are evaluated via HitRatio@1~\cite{liao2024llara} for identifying the correct item from candidate sets. For pairwise task, HitRatio@1 equals AUC (Area Under the Receiver Operating Characteristic)~\cite{bradley1997use}. We also adopt Valid Ratio to quantify valid response proportions across instances, assessing models’ instruction-following capability. Both metrics ensure response relevance and accuracy.

\subsubsection{Baselines}
We compared our PIDLR  with representative traditional and LLM-based recommendation methods.

\noindent\textit{\textbf{Traditional Recommendation Models}}

(1) \textit{Sequential recommendation} focuses on capturing temporal patterns in user-item interaction sequences to predict users' next interaction behavior. \textbf{GRU4Rec}~\cite{hidasi2015session} uses a Gated Recurrent Unit (GRU) to capture sequential patterns in user historical interactions.
\textbf{Caser}~\cite{tang2018personalized} adopts horizontal and vertical convolutions to capture high-order item sequence interactions for better recommendation. 
\textbf{SASRec}~\cite{kang2018self} integrates multi-head self-attention to effectively capture complex patterns in historical interaction data.

(2) \textit{Matrix factorization-based recommendation} decomposes user-item interaction matrices to capture latent factors for recommendation. \textbf{NeuMF}~\cite{he2017neural} combines linear matrix factorization with nonlinear MLP to extract low- and high-order features respectively. \textbf{LightGCN}~\cite{he2020lightgcn} streamlines GCNs for efficient user-item message propagation, simplifying architecture while preserving effectiveness. \textbf{SimpleX}~\cite{mao2021simplex} is a simple yet effective baseline integrating matrix factorization and user behavior modeling, optimized via contrastive loss and large-scale negative sampling.

(3) \textit{Knowledge graph-based recommendation} leverages auxiliary knowledge to enhance recommendation performance.
\textbf{KGAT}~\cite{wang2019kgat} employs attentive message passing on a knowledge-aware collaborative graph to effectively fuse node embeddings. \textbf{HGT}~\cite{hu2020heterogeneous} designs edge- and node-type dependent parameters for representation learning on heterogeneous graphs. \textbf{KGCL}~\cite{yang2022knowledge} applies contrastive learning to knowledge graphs, which helps reduce noise and guide user preference learning.

\noindent\textit{\textbf{LLM-based Recommendation Models}}

LLM-based recommendation methods leverage the powerful language understanding capabilities of LLMs to boost recommendation performance. 
\textbf{LLaMA3}~\cite{grattafiori2024llama} uses vanilla LLaMA3-8B Instruct with the same prompt as our work to generate recommendations.
\textbf{MoRec}~\cite{yuan2023go} enhances traditional recommenders by leveraging item modality features instead of sparse item IDs, integrating a text encoder and SASRec as the recommendation backbone.
\textbf{TallRec}~\cite{bao2023tallrec} pioneers LLM fine-tuning for recommendations by converting interaction sequences into textual prompts.
\textbf{LLaRA}~\cite{liao2024llara} introduces a hybrid item representation that combines traditional recommendation embeddings with textual tokens. This hybrid representation is then integrated into the fine-tuning process of the LLM.
\textbf{CoLaKG}~\cite{cui2024comprehending} leverages LLMs to comprehend knowledge graphs and generate semantic-enhanced representations for users/items, which are used to augment traditional recommendation.
\textbf{GLRec}~\cite{wu2024exploring} enhances recommendations by adopting a meta-path prompt constructor to improve heterogeneous knowledge graph comprehension of LLMs.
\subsubsection{Implementation Details}
All experiments were conducted on a single NVIDIA A100 GPU. The specific parameter settings for all experiments (including baselines) are detailed as follows. For Traditional Recommendation Models: the Adam optimizer was uniformly adopted with a learning rate of 0.001, an embedding dimension of 64, and a batch size of 256, plus early stopping on the validation set to avoid overfitting. These settings are consistent with those reported in prior studies~\cite{liao2024llara,bao2023tallrec}. Our PIDLR framework includes a traditional preference hint discovery stage, which leverages conventional models to extract crucial user attribute preferences and key attributes of items. To ensure fair comparison with conventional models, identical parameter settings are adopted for this stage. In contrast, the remaining parameter settings of PIDLR are detailed below: (1) For the pairwise task, the impact of core hyperparameters $N$, $\alpha_u$ and $\alpha_v$ is studied in Section~\ref{sec:hyper_parameter}, with optimal values determined experimentally. (2) For the listwise task, input text is lengthy (due to many candidate items and required attribute inclusion), highlighting the need to filter key attributes (a key contribution of this study). To balance performance and efficiency, $N=3$, $\alpha_u=0.1$ and $\alpha_v=0.3$ are set in Section~\ref{sec:hint discovery}. For LLM-based Models: LLaMA-3-8B Instruct~\cite{grattafiori2024llama} is utilized as the backbone model. Constrained by computation and for training efficiency, training is set to 5 epochs with a batch size of 128. To ensure fairness and consistency, MoRec baseline employs mean pooling of the item name's encoded embedding with LLaMA-3-8B Instruct to obtain the initial node embedding.

\subsection{Overall Performance Comparison (RQ1)}\label{sec:overall_performance}

\begin{table*}[]\small
    \centering
    \caption{
  Overall performance on pair-wise and list-wise  tasks. \textit{Impr\%} indicates the relative improvements of the best-performing method(bolded) over the strongest baselines(underlined).  Marker * indicates that paired t-test with p-value < 0.05.}\label{tab:performance} 
    \begin{tabular}{c|c||cc|cc||cc|cc}
    \toprule
    &&\multicolumn{4}{c||}{\textbf{Pair-wise Recommendation}} &\multicolumn{4}{c}{\textbf{List-wise Recommendation}} \\
    \cline{3-10}
         &\multirow{2}{*}{Model} &\multicolumn{2}{c|}{MovieLens} &\multicolumn{2}{c||}{LastFM} &\multicolumn{2}{c|}{MovieLens} &\multicolumn{2}{c}{LastFM}  \\
         \cline{3-10}
         &&Hit Ratio@1& ValidRatio  &Hit Ratio@1& ValidRatio &Hit Ratio@1& ValidRatio  &Hit Ratio@1& ValidRatio \\
         \hline
         \multirow{9}{*}{Traditional}& \textbf{GRU4Rec} &0.7902& 1.0000& 0.7519&1.0000  &0.2571& 1.0000& 0.2809&1.0000\\
         
         &\textbf{Caser} &0.7831 &1.0000 &0.7421&1.0000 &0.2509 &1.0000 &0.2738 &1.0000 \\
         
         &\textbf{SASRec} &0.7867& 1.0000&0.7059 &1.0000 &0.2563& 1.0000&0.2208&1.0000\\
         \cline{2-10}
         &\textbf{NeuMF} &0.7825& 1.0000&0.7228& 1.0000 &0.2393& 1.0000& 0.2717& 1.0000\\
         & \textbf{LightGCN} &0.7263& 1.0000& 0.7691 &1.0000&0.2306& 1.0000&0.2743 &1.0000\\
         & \textbf{SimpleX}&0.7803& 1.0000&0.7716 &1.0000&0.2368& 1.0000&0.2618 &1.0000\\
         \cline{2-10}
         
         & \textbf{KGAT} &0.7372 & 1.0000&0.7589&1.0000& 0.2581 & 1.0000&0.2918&1.0000\\
         & \textbf{HGT}  & 0.7827  &1.0000&0.7724&1.0000&0.2593& 1.0000&0.2814 &1.0000\\ %其实都还没跑完 0.7431& 0.7558 
         & \textbf{KGCL} &0.7831& 1.0000&0.7558 &1.0000&0.2406& 1.0000&0.2742 &1.0000 \\
         \hline
        \multirow{7}{*}{LLM-based}&\textbf{LLaMA-3}& 0.5142&0.0148&0.3333& 0.0053& 0.0684&0.4254&0.0539& 0.8202\\
        &\textbf{MoRec} & 0.7936 &1.0000& 0.7608&1.0000 & 0.2520 &1.0000& 0.3163&1.0000  \\ %0.8008
        &\textbf{TallRec} &0.7907 & 0.9966&0.7697& 0.9971&0.2627 & 0.9844&0.3971& 0.9898\\
        
        &\textbf{LLaRA} &0.7956 &0.9968 &\underline{0.7735}&0.9961&0.2727 &0.9875 &0.4002&0.9917  \\
        &\textbf{CoLaKG} &0.7988 &1.0000 & 0.7729 &1.0000 &0.2818 &1.0000 & 0.3987 &1.0000 \\
        &\textbf{GLRec} &\underline{0.7993} &0.9942 &0.7726&0.9945 &\underline{0.2821} &0.9810 &\underline{0.4073}&0.9896 \\
        \cline{2-10}
        &\textbf{PIDLR} &\textbf{0.8234}*&0.9937 &\textbf{0.8028}*&0.9967&\textbf{0.3012}*&0.9823 &\textbf{0.4235}*&0.9914 \\
        &\textbf{Impr\%} &3.0151 & - & 3.7880 &- &6.7706 & - & 3.9774 &-\\

        \bottomrule
        
    \end{tabular}

\end{table*}
Experimental results for pair-wise and list-wise tasks on both datasets are shown in Table~\ref{tab:performance}, with key observations as follows:
\begin{itemize}[leftmargin=*,itemsep=2pt,topsep=0pt,parsep=0pt]
    \item PIDLR significantly outperforms all baselines across pair-wise and list-wise tasks. For pair-wise recommendation, it achieves the highest HitRatio@1 (0.8234 on MovieLens, 0.8028 on LastFM), outperforming the best baseline by over 3.02\%. For list-wise tasks, its HitRatio@1 is 0.3012 (MovieLens) and 0.4235 (LastFM) with over 3.98\% improvements, confirming consistent superiority in recommendation accuracy.
    \item When evaluating conventional recommendation methods, we can analyze them from two key perspectives: (1) Most traditional knowledge-aware models (e.g. HGT) outperform non-knowledge-integrated models, as auxiliary knowledge can
    enhance user preference understanding;   (2)  These models still lag behind PIDLR, since PIDLR leverages LLMs’ advanced reasoning capabilities via learned preference hints.
    \item For LLM-based methods, three perspectives are further analyzed as follows: (1) Vanilla LLaMA3, which uses the same prompts as PIDLR but lacks fine-tuning, exhibits poor performance. This highlights the necessity of adapting LLMs specifically for recommendation tasks to enhance their effectiveness. (2) GLRec and PIDLR both integrate attribute-level knowledge into LLMs and outperform the LLM4Rec methods (e.g., MoRec, TallRec) without attribute knowledge, highlighting the importance of integrating the interacted attributes for recommendation. (3) GLRec underperforms PIDLR as interaction sparsity and excessive attribute noise limit its path-based knowledge graph methods in capturing user preferences for unseen items.

\end{itemize}
\subsection{Ablation Study (RQ2)}\label{sec:core_component}
As the \textbf{C}ollaborative preference h\textbf{I}nt \textbf{E}xtraction(CIE) and \textbf{I}nstance-wise \textbf{P}reference hint \textbf{D}iscovery(IPD) modules serve as the core modules of our model, we conduct the following ablation studies to verify their effectiveness.
\begin{itemize}[leftmargin=*,itemsep=2pt,topsep=0pt,parsep=0pt]   
    \item PIDLR(w/o IPD) excludes the discovered preference hints from input prompts, omitting attribute preference information from Eq. \ref{eq:user_attr} and Eq. \ref{eq:item_attr} in instance-wise preference discovery module.
    \item PIDLR(w/o CIE) removes the collaborative preference hint extraction module, deriving user attribute preferences directly from historical interactions, i.e., replacing  $\hat{\Gamma}_u$ with $\Gamma_u$  in Eq.\ref{eq:user_attr_p}.
    \item PIDLR(RD) randomly selects the same proportion of attributes for LLMs instead of sampling via scores from Eq.\ref{eq:user_attr_p} and Eq.\ref{eq:item_attr_p},  to evaluate the efficacy of our attribute hint selection strategy. 
    \item PIDLR(ALL) inputs all attributes into LLMs without any selection  i.e., replacing $\widetilde{\Gamma}_u$ with $\hat{\Gamma}_u$ in Eq. \ref{eq:user_attr} and  $\widetilde{\Gamma}_v$ with $\Gamma_v$ in Eq. \ref{eq:item_attr}. We initially designed this ablation experiment to incorporate all attributes (without selection) for list-wise recommendation. However, including all attributes in LLM prompts drastically prolongs training time and easily exceeds token limits, highlighting the \textbf{necessity of identifying key preference hints}.
    
\end{itemize}
From the results of ablation studies shown in Figure~\ref{fig:ab_test}, we can observe that:

\begin{figure}
    \centering
    \subfigure[Pair-wise recommendation task]{
    \includegraphics[width=0.45\linewidth]{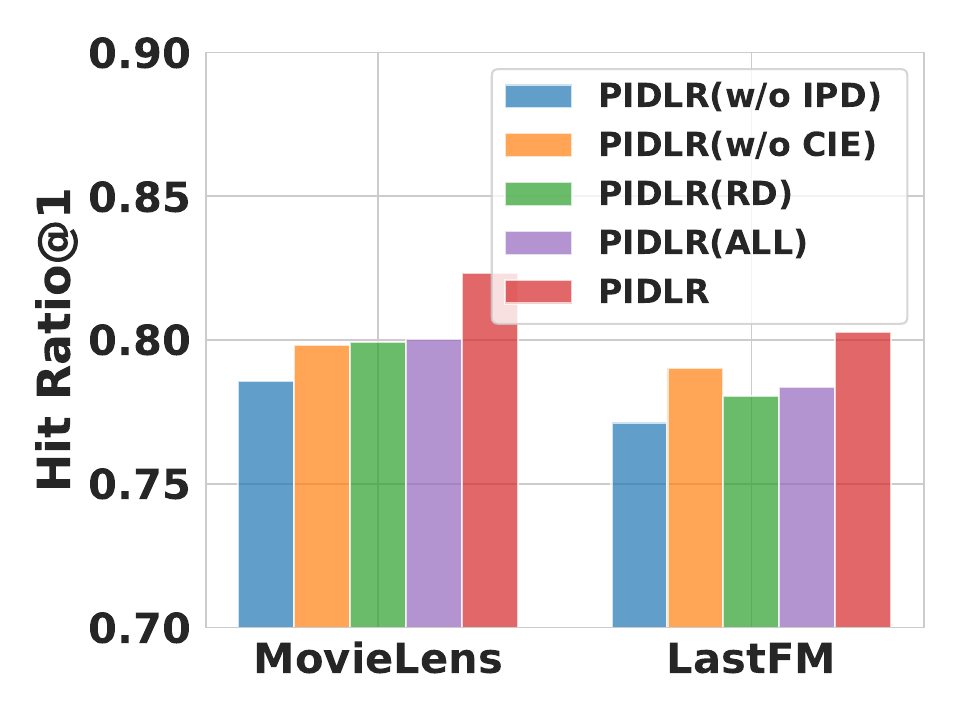}\label{fig:num_pair}
    }
    \subfigure[List-wise recommendation task]{
    \includegraphics[width=0.45\linewidth]{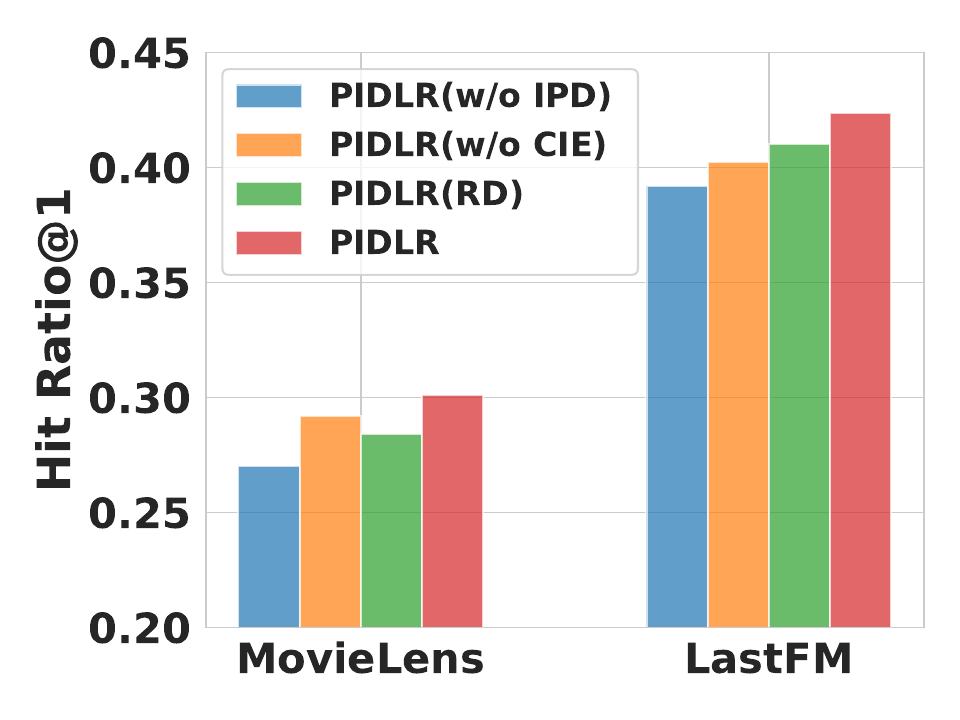}
    }
    \caption{Ablation study on collaborative preference hint extraction and instance-wise preference discovery modules.}
    \label{fig:ab_test}
\end{figure}

\begin{itemize}[leftmargin=*,itemsep=2pt,topsep=0pt,parsep=0pt]
    
    \item PIDLR outperforms PIDLR(wo IPD), which removes the instance-wise preference discovery, as additional attribute information provides auxiliary evidence for recommendations.
    \item PIDLR outperforms PIDLR(wo CIE), which lacks the collaborative preference hint extraction module, across both datasets. This improvement stems from the module’s ability to supplement users’ potential preferences for unseen items, which may not be adequately captured by their sparse interactions.
    \item PIDLR surpasses both PIDLR(RD) and PIDLR(ALL), which randomly select attributes and input all attributes, respectively, underscoring the critical role of instance-wise discovery in enabling LLMs to identify core preference hints for each instance.

\end{itemize}

\subsection{Case Studies (RQ3)}
\begin{figure*}
    \centering
    \subfigure{
    \includegraphics[width=0.6\linewidth]{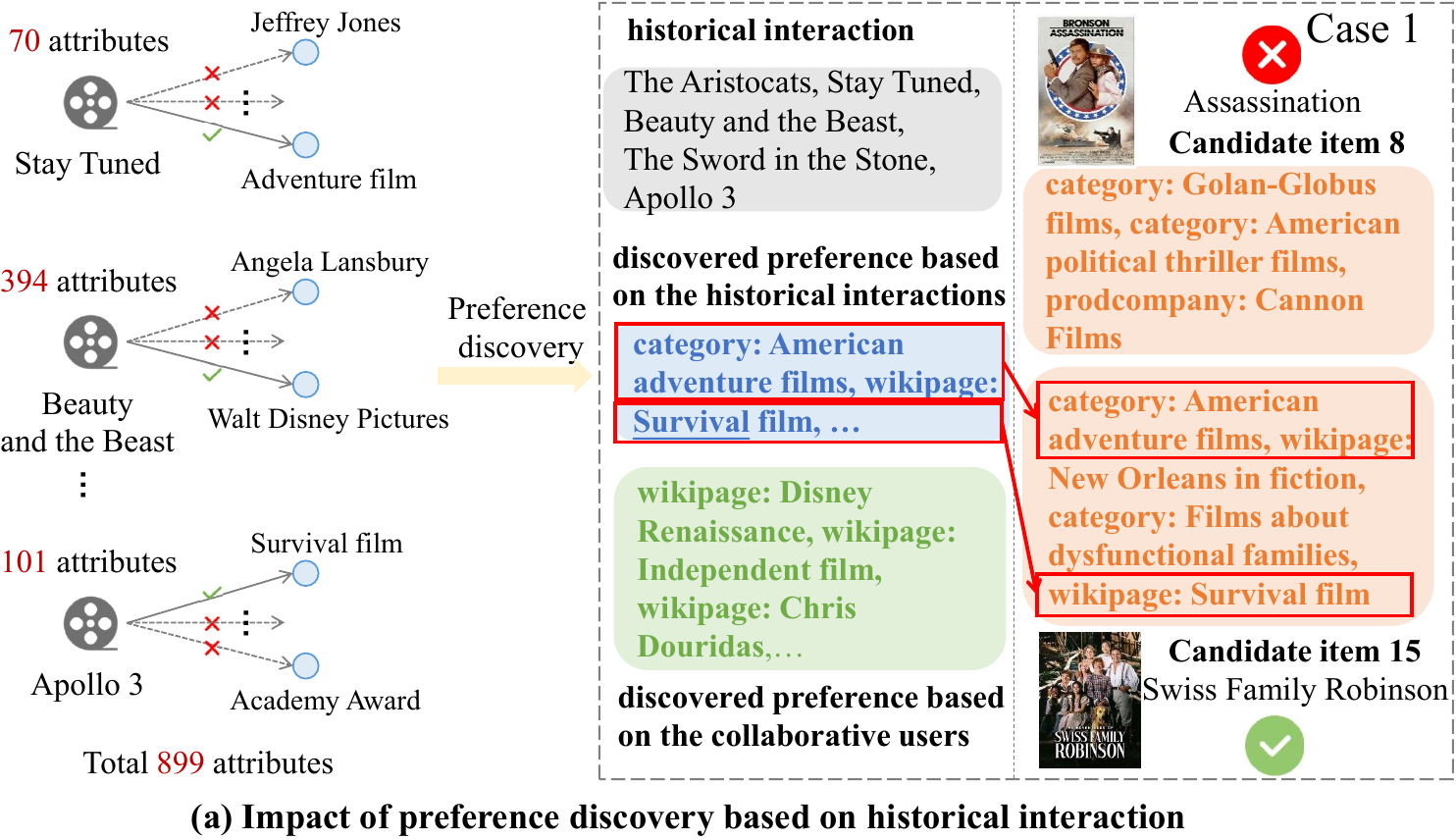}
    }
    \subfigure{
    \includegraphics[width=0.348\linewidth]{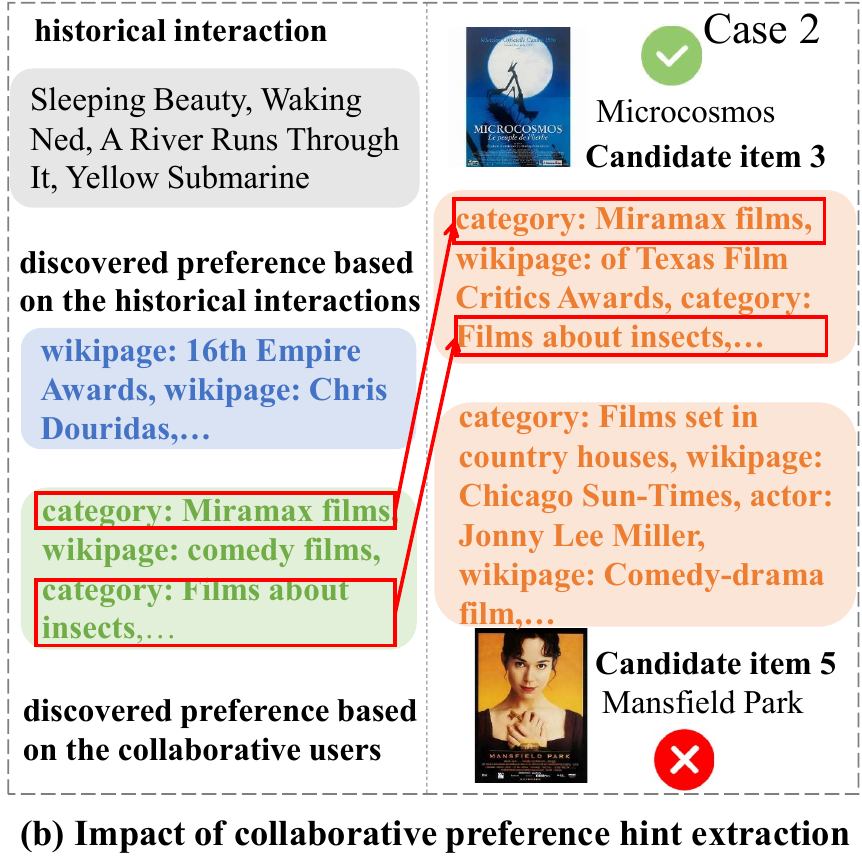}
    }
    \caption{Case Studies. (a) PIDLR infers preference hints for \textit{adventure/survival films} from the user’s historical interactions in the left-side knowledge graph, achieving successful recommendation of \textit{Swiss Family Robinson}. (b) Collaborative preference extraction reveals potential interest in \textit{Miramax/insect films}, enabling accurate recommendation of \textit{Microcosmos}.}\label{fig:case_study}
\end{figure*}

Our PIDLR aims to discover instance-specific essential preference hints and integrate them into LLMs, thereby enhancing preference hint comprehension for recommendation. To verify LLMs’ capacity to leverage these hints for accurate recommendations, we conducted case studies with two representative samples as shown in Figure~\ref{fig:case_study}.
In Case 1, PIDLR infers a user’s potential preference for American adventure films and survival films from historical interactions. The candidate \textit{Swiss Family Robinson} aligns more closely with these hints than \textit{Assassination}, enabling the LLM to effectively recommend the ground-truth candidate. This demonstrates our model’s strong \textbf{ability to identify critical preference hints}.
In Case 2, preference hints derived solely from historical interactions are insufficient to distinguish between candidates \textit{Microcosmos} and \textit{Mansfield Park}. By incorporating collaborative user data, the model reveals that similar users favor \textit{Miramax} films and insect-related films, attributes that \textit{Microcosmos} better matches. These insights equip the LLM with robust hints to select the ground-truth \textit{Microcosmos}, highlighting \textbf{the collaborative preference hint extraction module’s significance in capturing comprehensive user preferences and improving recommendation accuracy}.
\subsection{Impact of Hyperparameter(RQ4)}\label{sec:hyper_parameter}
We conducted experiments on the LastFM dataset to explore the impact of critical hyperparameters on the pairwise recommendation task. Specifically, we investigated three key hyperparameters: (1) $N$ in the Collaborative Preference Hint Extraction module (Section~\ref{sec:coll_dis}); (2) $\alpha_u$ for user preference discovery (Section~\ref{sec:User_Preference_Discovery}); and (3) $\alpha_v$ for item attribute discovery in the Instance-wise Hint Discovery module (Section~\ref{sec:item_Attribute_Discovery}).

\begin{figure}
    \centering
    \subfigure[the number of collaborative users $N$]{
    \includegraphics[width=0.3\linewidth]{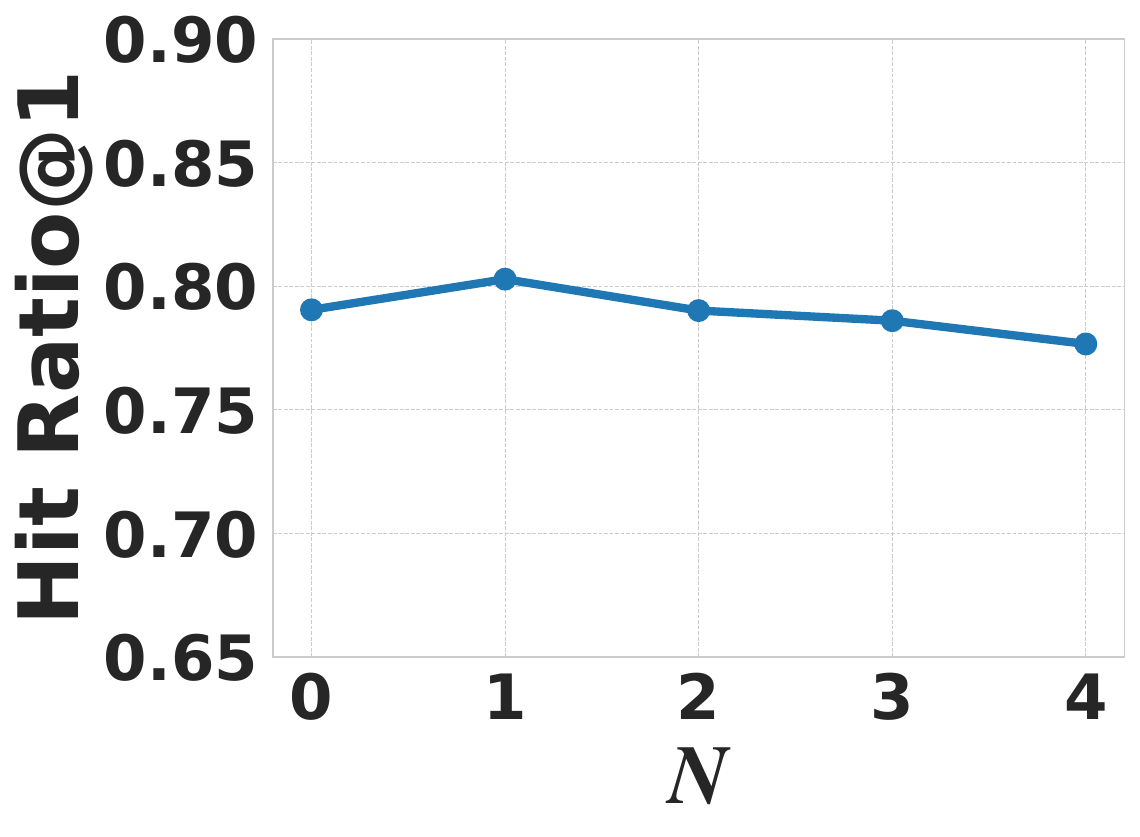}\label{fig:collaborative_num}
    }
    \subfigure[selection ratio of user preference hints $\alpha_u$]{
    \includegraphics[width=0.3\linewidth]{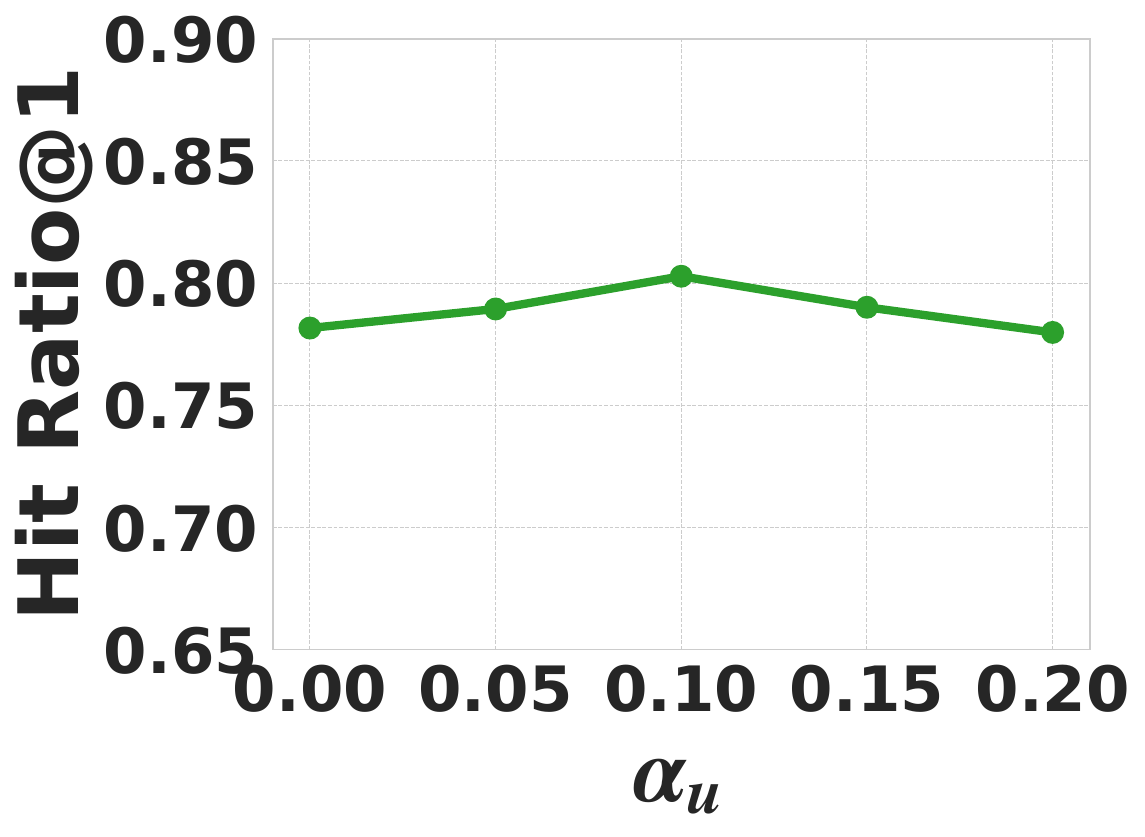}\label{fig:user_ratio}
    } 
    \subfigure[selection ratio of item attribute hints $\alpha_v$]{
    \includegraphics[width=0.3\linewidth]{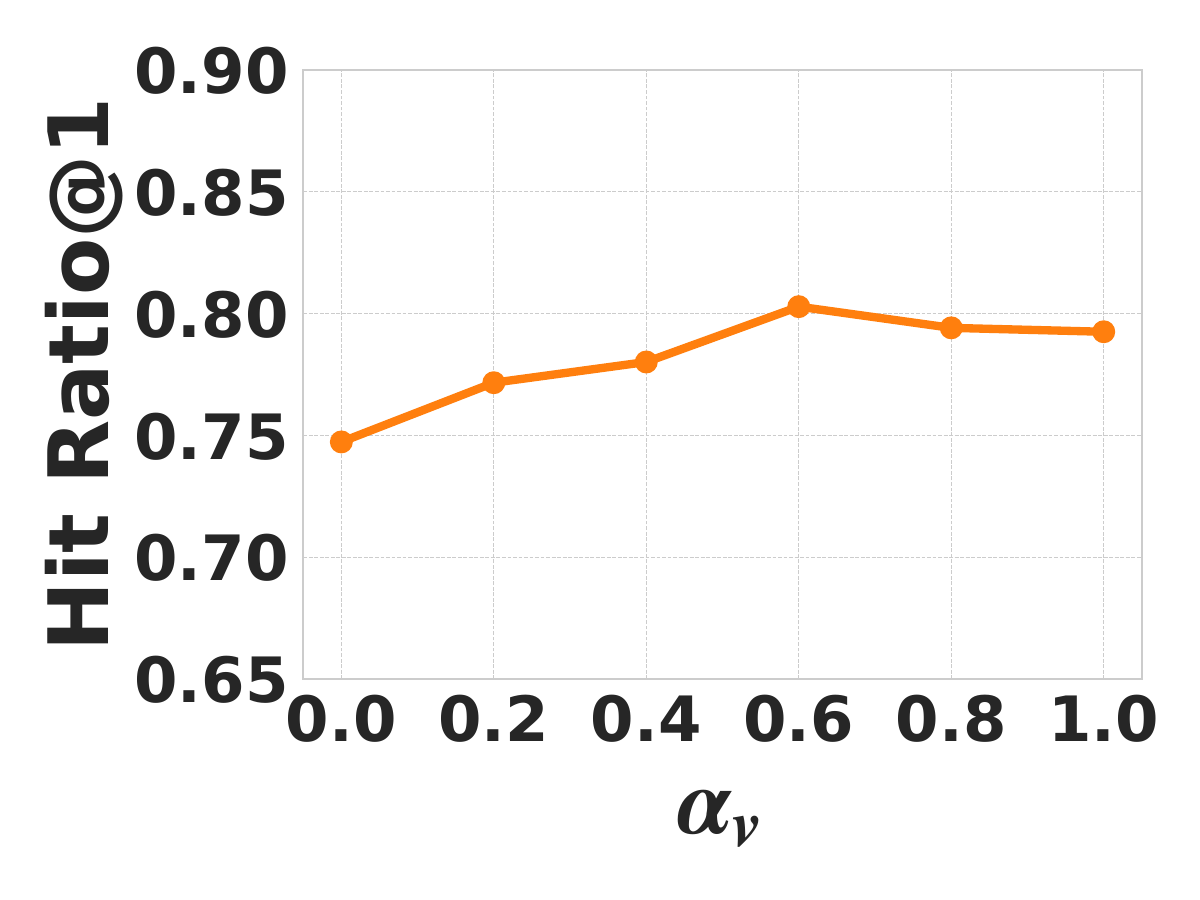}\label{fig:item_ratio}
    }
    \caption{Impact of Hyperparameter.}
    \label{fig:hyper_test}
\end{figure}

\noindent\textbf{(1) Impact of Collaborative Preference Hint Extraction.}
To investigate the impact of the number of collaborative users, we vary $N$ in $[0, 1, 2, 3, 4]$ (Figure~\ref{fig:collaborative_num}). Performance first rises with increasing $N$ since integrating collaborative filtering into LLMs can enhance recommendation hint understanding. This finding indicates sparse interactions alone cannot reflect unseen-item preferences.
However, excessive $N$ introduces low-similarity user noise, degrading performance. Overall, the results validate the effectiveness of explicit collaborative information integration into LLMs.

\noindent\textbf{(2) Impact of User Preference Discovery.}
To investigate the impact of the selection ratio of user attribute-level preferences, we vary $\alpha_u$ in $[0, 0.05, 0.10, 0.15, 0.20]$, with results shown in Figure~\ref{fig:user_ratio}. Performance improves then declines as $\alpha_u$ increases.
When $\alpha_u = 0$, no user preference information is input into the LLM. Initial gains stem from crucial preference information that boosts LLM preference understanding, with optimal performance at $\alpha_u=0.10$. Excessive $\alpha_u$ (e.g., 0.20) degrades performance even below $\alpha_u=0$, indicating \textbf{blindly incorporating excessive user-item attribute information introduces severe noise}. These results underscore the necessity of identifying true user preferences and selectively feeding them into LLMs to enhance recommendation performance.

\noindent\textbf{(3) Impact of Item Attribute Discovery.}
To study the impact of selection ratio of item attributes, we vary $\alpha_v$ in $[0.0, 0.2, 0.4, 0.6, 0.8, 1.0]$, with results illustrated in Figure~\ref{fig:item_ratio}. The result shows that the model's performance initially improves and then declines as $\alpha_v$ increases. $\alpha_v = 0$ means no item attribute input, leading to poor performance. Initial gains come from key attribute information that enhances LLM instance-specific item understanding, with optimal performance at $\alpha_v=0.60$ (i.e., the 60\% most relevant attributes). As $\alpha_v$ continues to increase, the performance declines, as \textbf{including irrelevant attributes introduces model noise and impairs effectiveness}. These results validate the crucial role of  this module in filtering noise and extracting key factors to boost LLM-based recommendation performance.

\subsection{Model Efficiency in Few-shot Scenario (RQ5)}

\begin{figure}
    \centering
    \subfigure[Pair-wise task ]{
    \includegraphics[width=0.45\linewidth]{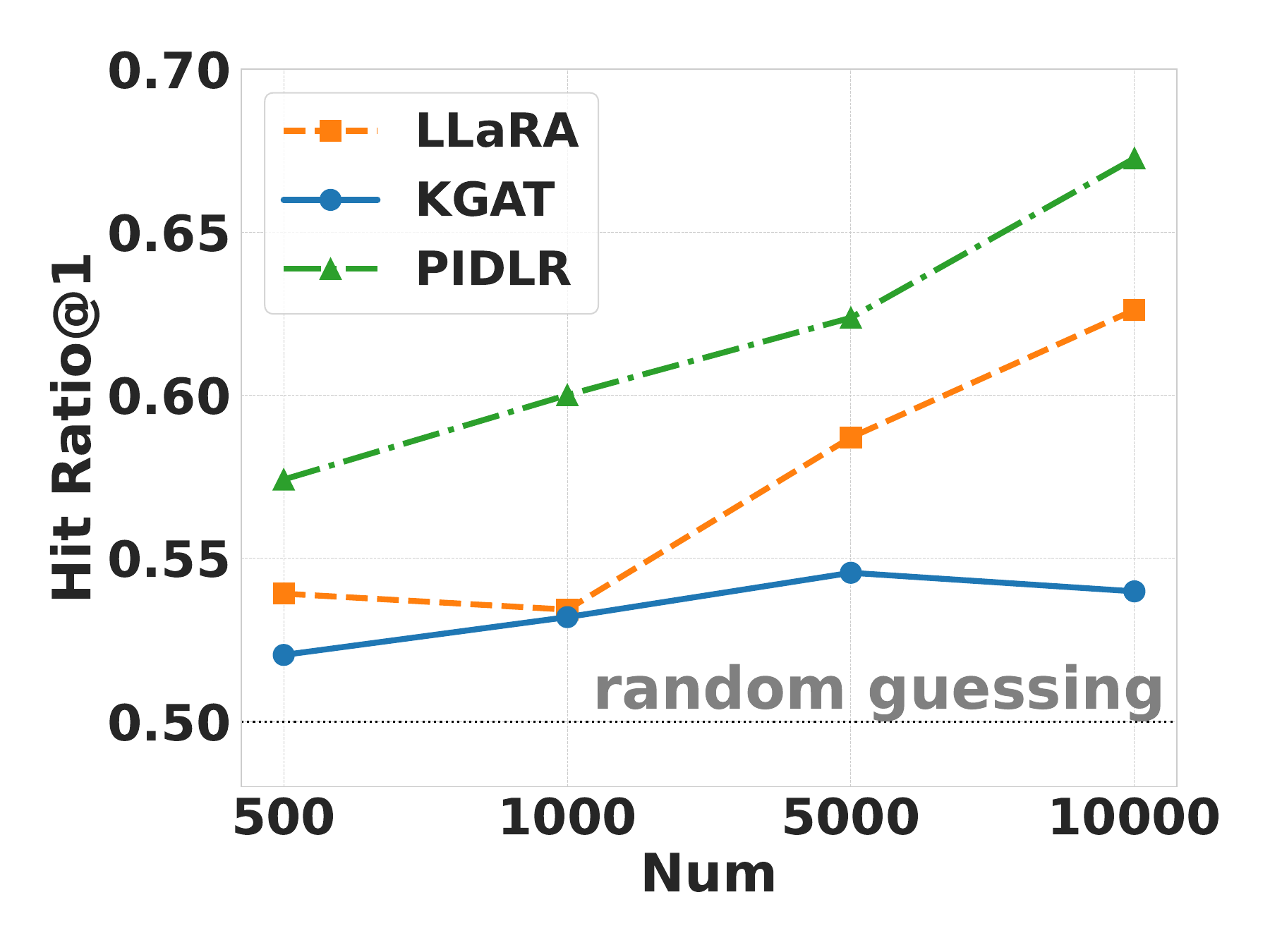}\label{fig:num_pair}
    }
    \subfigure[List-wise task]{
    \includegraphics[width=0.45\linewidth]{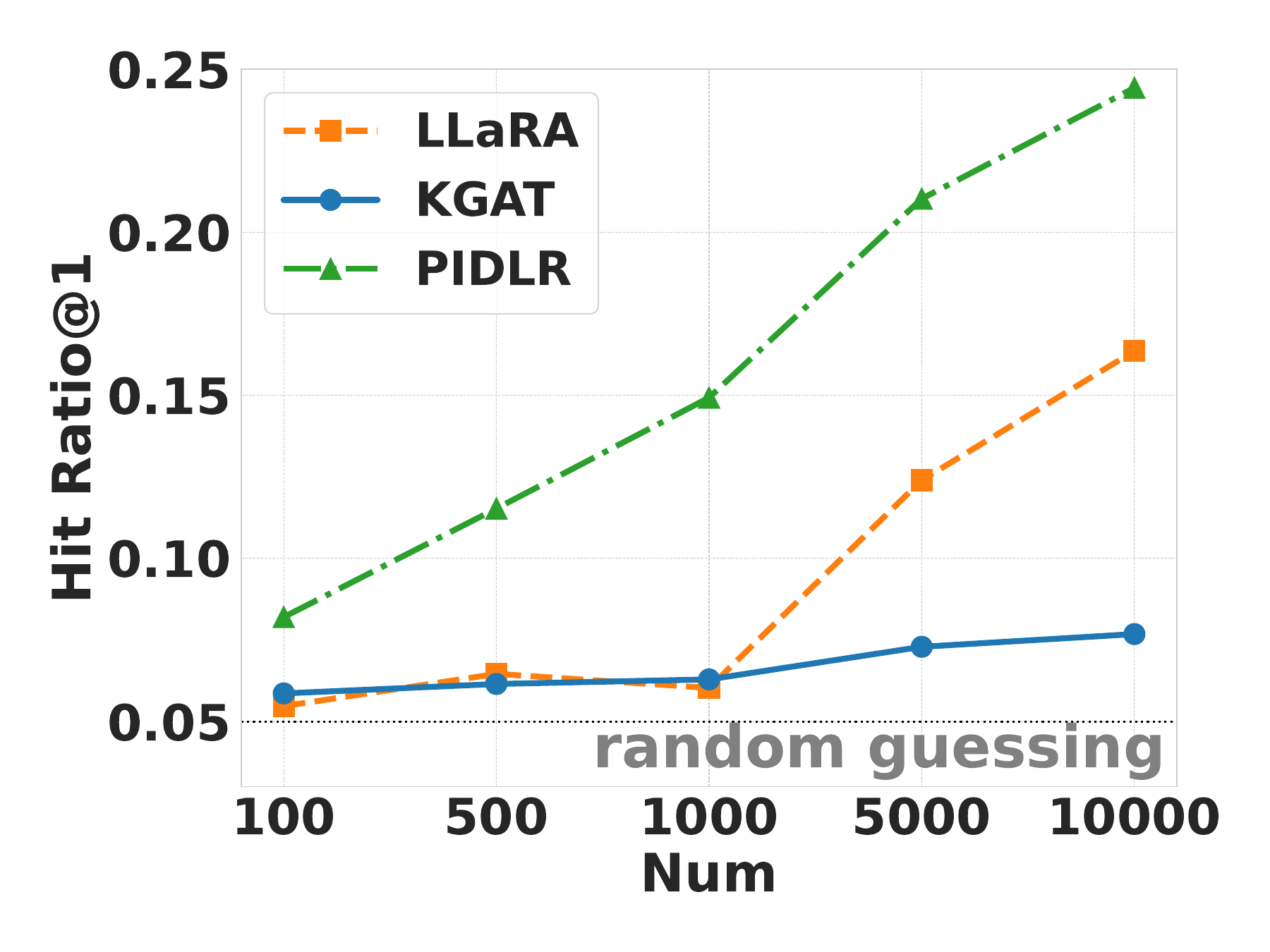}\label{fig:num_list}
    }
    \caption{Performance w.r.t Number of Training Instances ($Num$)  on LastFM. Random guessing yields Hit Ratio@1=0.5 and 0.05, with 2 and 20 candidates, respectively.}
    \label{fig:performance_num}
\end{figure}

The few-shot learning capability is a critical metric for LLMs, as it evaluates whether the model grasps the core logic of the recommendation task. Here, we assess the model’s few-shot performance by investigating how variations in training sample size affect overall recommendation performance. With limited training data, models tend to encounter numerous unseen items during inference, which impairs their ability to generate accurate recommendations. We compare our model against two baselines: the knowledge-aware KGAT and the LLM-based LLaRA. Figure~\ref{fig:performance_num} shows model performance across different training sample sizes for both pair-wise and list-wise tasks on the LastFM dataset. Performance generally improves with training data size across tasks. Our model consistently outperforms the LLM-based baseline LLaRA across all sample sizes. This advantage comes from its enhanced ability to understand recommendation-oriented preferences, which enables better adaptation to limited training data. Besides, it also surpasses the knowledge-based baseline KGAT across all sample sizes. This advantage is largely driven by the superior reasoning capabilities of LLMs, which allow our model to leverage discovered preference hints more effectively. Notably, when using small datasets (i.e., $\text{Num} \leq 1000$), LLaRA and KGAT perform near random guessing, especially in the more challenging list-wise task with more candidates. In contrast, our model maintains reasonable performance on both tasks with minimal data. Overall, integrating preference hints into LLMs enables better understanding of the core recommendation logic and boost LLM-based recommendation performance.%recommendations under limited data, highlighting our model’s strong few-shot generalization and adaptability.

\section{Conclusion}\label{sec:conclusion}
This paper introduces PIDLR, a novel model that uncovers instance-wise preference cues based on traditional recommendation principles, thereby guiding LLMs to grasp recommendation rationales. Specifically, we first propose a collaborative preference hint extraction module to identify similar users and supplement potential user preferences for unseen items. To mitigate noise from extensive attributes, an instance-wise preference discovery module is introduced to identify preference hints specific to each instance. The identified hints are then converted into head-centric flattened prompts for LLMs. Empirical results show PIDLR outperforms traditional and LLM-based baselines. Further few-shot experiments verify our model helps LLMs better understand recommendation logic. Notably, PIDLR extracts semantic hints as prompts via \textbf{traditional recommendation principles}. Leveraging the computational efficiency of traditional models, we select key feature sets that help LLMs comprehend tasks and simplify prompts. This paradigm also has broader applicability: traditional methods can pinpoint essential factors to guide LLMs in large feature-set scenarios.
% unlike existing methods that generally embed implicit, id-level user/item vectors (\textbf{mismatched with LLM semantic space}) into LLMs, In the future, we also plan to explore model-free approaches. Specifically, we will use post-hoc Shapley values to discover the most important factors without relying on specific  model structures.

% \begin{acks}
% This work is partly supported by the National Key Research and Development Program of China (No. 2023YFF0725001), the National Natural Science Foundation of China (No.62306255, 92370204, 62176014), the Natural Science Foundation of Guangdong Province (No. 2024A1515011839), the Fundamental Research Project of Guangzhou (No.
% 2024A04J4233), the Guangzhou-HKUST(GZ) Joint Funding
% Program (No.2023A03J0008), and the Education Bureau of Guangzhou Municipality.
% \end{acks}

\bibliographystyle{ACM-Reference-Format}
\bibliography{ref}
\clearpage
\end{document}